\documentclass[12pt]{article}

\usepackage[screen]{geometry}
\usepackage{palatino}
\usepackage[utf8]{inputenc}
\usepackage{amsmath,amssymb}
\usepackage{latexsym}
\usepackage{circle}
\usepackage{parskip} 
\usepackage{hyperref}
\usepackage{graphicx}
\usepackage{forest}
\usepackage{mathtools}
\usepackage{proof}

\allowdisplaybreaks 

\usepackage{mdframed}

\newcommand{\node}[1]{\textsf{#1}}

\newcommand{\doi}[1]{\href{http://dx.doi.org/#1}{\textsf{doi:#1}}}

\newcommand{\bLozenge}{\mathbin{\blacklozenge}}

\begin{document}
\urlstyle{sf}

\title{Towards Deriving Verification Properties}
\author{Michael Winikoff}
\date{\today} 

\maketitle

\section{Introduction}

\begin{quote} 
``\textit{\textbf{There are two key problems that occur in all such approaches: (1) what properties do we verify}; and (2) where
do the probabilities come from. \textbf{The first of these remains a problem with all formal analysis techniques} and, clearly,
significant work must be done in capturing the requirements of the system in a formal and logical
way}.''~\cite{DBLP:journals/fac/KonurFDK14} (emphasis added) 
\end{quote}

Formal software verification uses mathematical techniques to establish that software has certain properties. For example, that
the behaviour of a software system $S$ satisfies certain logically-specified properties. A common approach is \emph{model
checking}~\cite{ModelCheckingBook} which takes a model of a system ($M$)) and a property $\phi$ (typically specified in some
temporal logic), and can return either a guarantee that the behaviour of $M$ satisfies $\phi$, or a counter-example: a possible
behaviour of $M$ that violates $\phi$.

Formal methods have a long history~\cite{Fetzer88:veryidea,dMLP79,BoyerMoore81} but, as suggested by the opening quote, a
recurring assumption is that the property $\phi$ is known, or provided as part of the requirements elicitation process.  

This working note considers the question: \emph{where does $\phi$ come from?} 

Our answer to this question is a pragmatic one: we define a \emph{process} that can be used to systematically identify a set of
verification properties. 

The high-level idea is that we start with a well-justified (and universal?) collection of generic high-level properties
(``tenets'') such as ``do no harm''. We then systematically (but informally) derive contextualised more specific properties,
using \textbf{domain knowledge} and elements of the \textbf{system's design}. These more specific properties capture the ways in
which the specific system can violate the desired high-level properties. 

For example, imagine a robot that assists an elderly person living in their home. One high-level property (``tenet'') might be
that the person should be kept safe (i.e.~safe from harm, and healthy). This high-level property clearly cannot be specified
formally. However, what we can do is carefully (and systematically) consider all the ways in which the person could come to
harm, given the context of the system and its functionality, along with domain knowledge. This might allow us to (informally)
derive specific properties, such as that the person is accompanied if they leave the house, and that they are reminded to take
medication.

\begin{figure}[ht]
\center
\includegraphics[width=\textwidth]{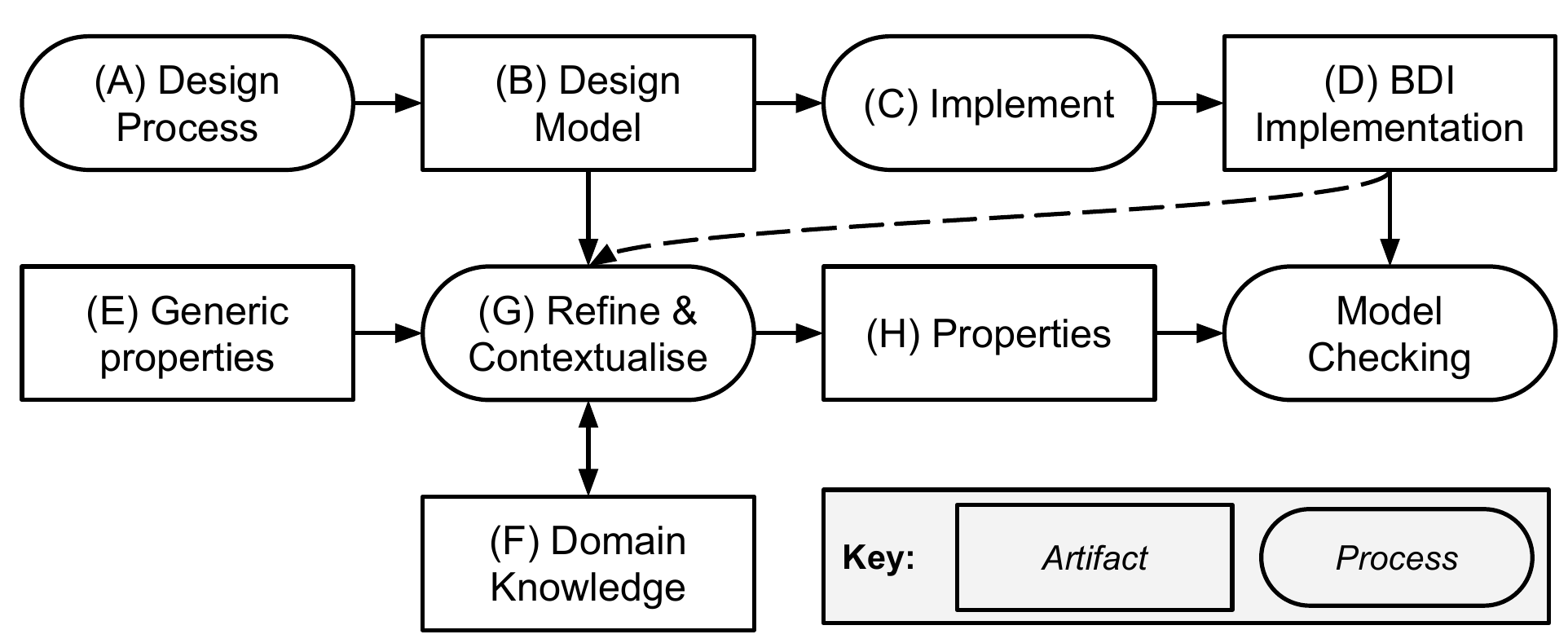}
\caption{High-level process}\label{fig:highlevel}
\end{figure}

We now define the high-level process in a little more detail. We assume (see Figure~\ref{fig:highlevel}) that a well-defined
design process (A) is followed, which results in some design models (B), and that these are then refined and implemented (C),
yielding code in some appropriate programming language, such as a BDI (Belief-Desire-Intention) agent-oriented programming
language (D). 

The process that we follow then takes as input high-level generic properties (``tenets'') (E), as well as domain knowledge (F),
and information about the design (B) and possibly even the implementation (D), and uses the domain and design information to
contextualise and refine (G) the high-level properties into more specific properties (H).  These specific properties and the
software can be model checked.  Note that this process proceeds from informal to formal. 

For the remainder of this document we make the following assumptions about the specific forms of these different artefacts.

For the design model (B) we remain as agnostic as possible, and only assume some form of goal model where goals are related to
their sub-goals. This sort of model is common to many AOSE methodologies~\cite{AOSEChapter}.

For the implementation (D) we actually do not need to make any assumptions, as long as (B) has a goal model. If it does not,
then, if the implementation uses a notion of goals, then we can extract a goal-tree from the implementation, shown as a dashed
line in Figure~\ref{fig:highlevel}.
    
For the tenets (E), we assume simply English text. We do not believe that it is possible to effectively formalise high-level
notions such as ``does no harm'', in fact, the whole point of the process we propose, is to allow such properties to be captured
formally by making them more specific in the context of the system at hand.
    
Domain knowledge (F)  could be represented in a number of ways. We assume (following~\cite{DBLP:journals/tse/LamsweerdeL00})
that it is represented as a collection of implications. However, we could also explore using a graphical model instead of
logical formulae. Note that domain knowledge is also refined and extended as part of the process (G), indicated by the
bi-directional arrow between (G) and (F).

Properties to be verified (H) are logical formulae. A wide range of logics could be used, but for the present paper we simply
assume Linear Temporal Logic (LTL). Other richer logics could be used, but there is the standard trade-off that the richer the
logic, the harder  verification becomes.
    
The remainder of this document provides a sketch for a possible process, illustrated using a running example. This process is an
early sketch, and is intended as a starting point for discussion and refinement.

Before proceeding further, it is worth briefly mentioning a key paper that informs this work. 

The process of deriving ways in which a tenet could be compromised has similarities with a 2000 paper by van Lamsweerde and
Letier (henceforth ``vLL'')~\cite{DBLP:journals/tse/LamsweerdeL00}. Very briefly, they begin with leaf goals (i.e.~requirements,
which are formalised in logic), and derive \emph{obstacles} to these requirements, using domain knowledge. 

Although the starting point is different, a formally specified system goal in their approach vs.~an informal generic tenet in
ours, the general idea of deriving a more detailed obstacle for the starting point, by using domain knowledge, is similar. Note
that sometimes the derivation of obstacles in their method uses not domain knowledge, but a general pattern that a goal $G$ can
go wrong by not being done at all, or by being done wrongly, where ``done wrongly'' can refer to any aspect of the goal.  So,
for example, if the goal is to send a message, then it could go wrong by the message not being sent, or it could be sent to the
wrong recipient, or sent with incorrect contents. 

One difference, is that whereas they start with a specific goal, we start with a high-level tenet. This is because the tenets
are what we care about and want to ensure, whereas not all goals are important. For example, in a home-care robot scenario,
checking that the fridge is not left open is not that important, whereas ensuring that the person being cared for is reminded to
take their medication is very important.

\section{Artefacts}

This section briefly describes the artefacts that play a role in the process.  As depicted in Figure~\ref{fig:highlevel}, the
process (G) involves tenets (E), a design model, specifically a goal tree (B), and domain knowledge (F). It results in formally
specified properties to be verified (H). 

We have already noted that the tenets (E) are simply high-level and generic statements in English, for example ``do not harm
humans'', or ``ensure the system is able to continue to function''. 

We have also already noted that the formulae to be verified (H) are specified in Linear Temporal Logic (LTL), but that other
logics are also possible.

\begin{figure}[ht]
\center
\includegraphics[width=\textwidth]{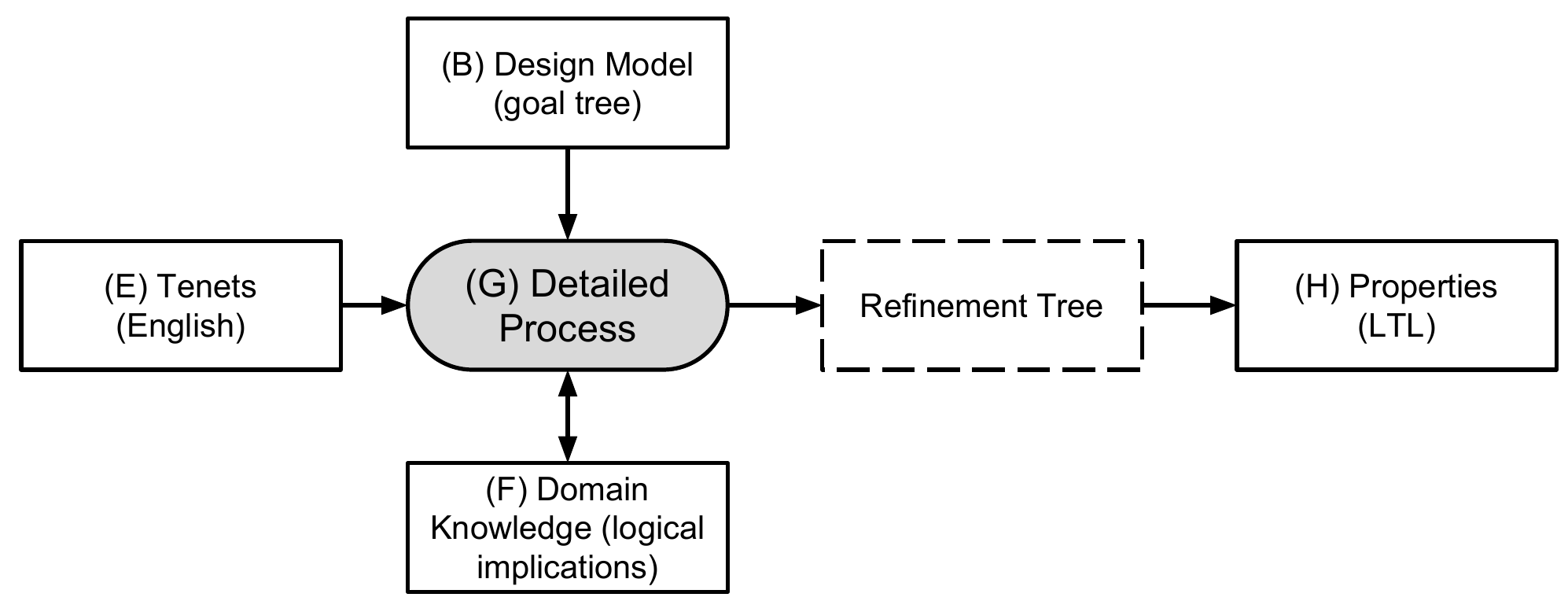}
\caption{Zooming in on step G}\label{fig:newprocesszoom}
\end{figure}

We now turn our attention to the goal tree (B) and domain knowledge (F). We also define an intermediate data structure: the
refinement tree (see Section~\ref{sec:refinementtree}). Figure~\ref{fig:newprocesszoom} shows the process, along with the key
artefacts.

\subsection{Goal Tree}

\begin{figure}[ht]
\centering
\includegraphics[width=\textwidth]{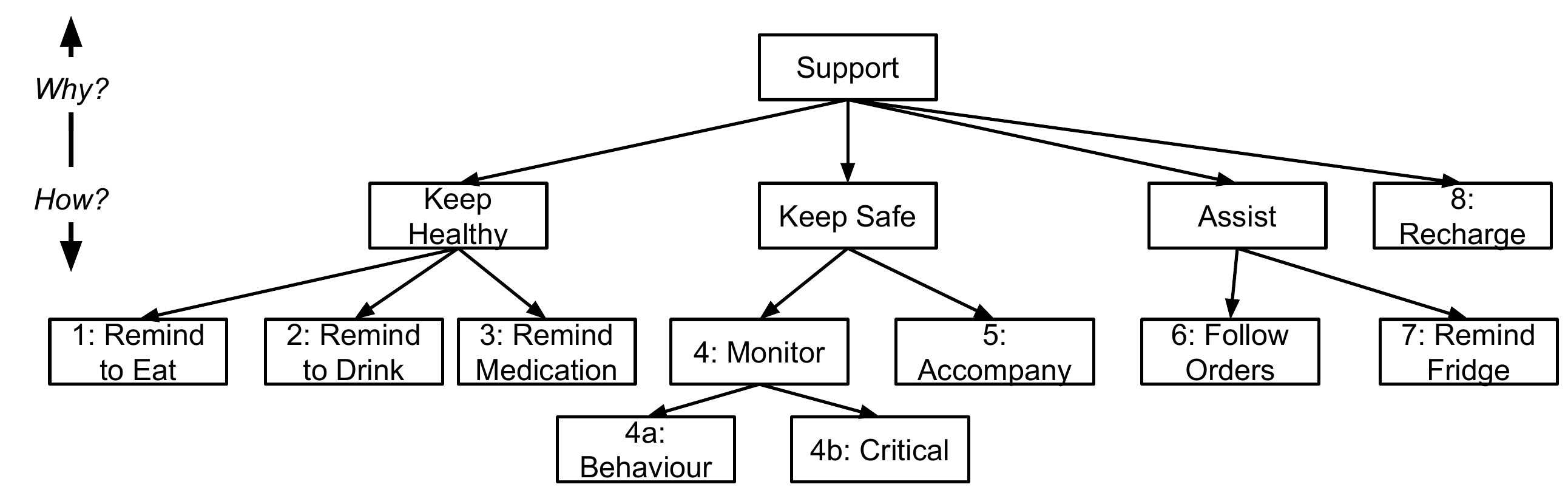}
\caption{Example Goal Tree}\label{fig:goaltree}
\end{figure}

The design model that we use is a \textbf{goal tree} (see example in Figure~\ref{fig:goaltree}). This is a tree\footnote{More
precisely a directed acyclic graph, since a sub-goal can be reused, i.e.~it can be the child of multiple parents.} where nodes
are (sub-)goals, and arrows link goals to their sub-goals. The relationship of a goal to its children is indicated as being
either ``AND'' or ``OR'' (in the example in Figure~\ref{fig:goaltree} all the relationships are ``AND''). Such trees can be
constructed and refined by asking ``Why?'' to identify parents of goals, and asking ``How?'' to identify children of
goals~\cite{lamsweerde01,PrometheusBook}. Such trees are commonly used in methodologies for engineering autonomous
systems~\cite{AOSEChapter}.  

The tree in Figure~\ref{fig:goaltree} relates to a Care-O-bot scenario~\cite{DBLP:conf/iros/ReiserCFKBWJPHV09}. This scenario
involves a robot in the home of an elderly person. The robot reminds the person to eat, drink, and take their medication, is
able to monitor them and alert medical authorities if required, and also performs a range of other support tasks (e.g.~checking
if the fridge door is left open, watching TV together, following orders to fetch items, answering the door).

\subsection{Domain Knowledge}

The third artefact is \textbf{domain knowledge} which we model as implications in a suitable logic (for example, LTL, following
vLL).

One example domain knowledge rule captures that issuing reminders (e.g.~to take medication) tends to lead to the task in
question being done.  This is actually difficult to formalise precisely, since the reminders do not actually guarantee
compliance. For simplicity we formalise the overly-strong property that reminding will ensure compliance: 
\begin{equation}
\Box (\mathit{remind}(X) \rightarrow \mathit{do}(X)) \label{eqnn1}
\end{equation}
Of course, this is too strong, and a more nuanced formalisation might involve a logic with probabilities, allowing one to
capture that reminders reduce the probability of forgetting, while not eliminating it entirely. 

Conversely, we may also know that reminders have value because the person being cared for is forgetful, and is more likely to
remember to perform a task if they are reminded. Again, a correct formalisation would require probabilities, but for now we
formalise using an overly strong property that reminders are necessary (i.e.~without reminders, the person will forget): 
\begin{equation}
\Box (\lnot \mathit{remind}(X) \rightarrow \lnot \mathit{do}(X)) \label{eqnn2}
\end{equation}
(of course, we could simply have written the earlier property using
$\leftrightarrow$)

Note: For convenience, we adopt the convention (used by vLL) that $A \Rightarrow B$ is notational shorthand for $\Box (A
\rightarrow B)$, which allows us to write the first domain knowledge above as simply $ \mathit{reminder}(X) \Rightarrow
\mathit{do}(X) $.

Other examples of domain knowledge include defining high-level concepts, e.g.~ that ``not getting enough food'' means having
fewer than three meals a day (and obviously there need to be conditions on what those meals are), and that ``not drinking
enough'' is less than around 1.2 litres per day. We use ``$\equiv$'' to define such equivalences, which logically is treated as
bidirectional implication.
\begin{eqnarray}
\mbox{not enough food} & \equiv & \mbox{$<$3 meals a day} \label{eqn3} \\
\mbox{not enough drink}& \equiv & \mbox{$<$1.2L/day} \label{eqn4} \\
\mbox{accompany excursion} & \equiv & \mbox{follow or delegate-by-informing} \label{eqn5} \\
\mbox{keep healthy} &\Rightarrow & \mbox{enough food} \land \mbox{enough drink} \land \mbox{correct medication} \label{eqn6} \\
\mbox{$<$1.2L/day} & \equiv &  \lnot (do(\mathit{drinkregularly})) \label{eqn10} \\
\mbox{correct medication} & \equiv & \mbox{issued $ = $ prescribed}\label{eqn8}  \\
\mbox{$<$3 meals a day} &\equiv &  \lnot (do(\mathit{breakfast}) \land do(\mathit{lunch}) \land do(\mathit{dinner})) \label{eqn9} 
\end{eqnarray}

Additionally, note that the goal tree can be seen as specifying implied domain knowledge, for instance if a goal $G_0$ has a
child goal $G_1$ then this indicates that $G_1$ is part of achieving $G_0$. However, in the detailed process presented below, we
keep the goal-tree distinct, rather than mapping it to domain knowledge rules.

\subsection{Refinement Tree}\label{sec:refinementtree}

The artefacts discussed so far are the inputs to the process. The process uses, and potentially modifies, these to progressively
generate a \textbf{refinement tree}. This is an intermediate artefact (hence shown dashed in Figure~\ref{fig:newprocesszoom})
that captures the process of refining the tenet.  In practice, this tree is valuable for traceability. However, it is not
required: we could equally well work with just a set of nodes (the frontier of the refinement tree).

The root of a refinement tree is a tenet, and, once the process is completed, its leaves are each an LTL property.  Each node in
the refinement tree is a description (formal or informal) of a \emph{set of behaviours}. That is, the node specifies a subset of
the possible behaviours of the system. The subset of behaviours is described in terms of a condition, which, since it can be
instantaneous, or over a time period, is formalised (eventually) in LTL. 

The relationship between a node and its children is \emph{implication}: if node $B$ is a child of node $A$, then that means that
any behaviour that satisfies $B$ will also satisfy $A$. So, for example, given a node ``\node{not enough food}'', with a child
node ``\node{$<$3 meals a day}'', this implication holds.  Since having less than 3 meals a day is (according to domain
knowledge Equation~\ref{eqn3}) the definition of not having enough food, any behaviour that meets the condition of not
having 3 meals a day, by definition also meets the condition of not having enough food.  Similarly, consider a node
``\node{harm}'' with two child nodes: ``\node{not kept healthy}'' and ``\node{put at risk}''. A behaviour that satisfies one of
these, by either not keeping the person healthy or by putting them at risk, also is considered to be meeting the definition of
causing them harm\footnote{Note that in this example we are considering putting a person at risk of coming to harm, in a
situation where the system cannot prevent the harm, as being equivalent to actually causing harm, even though the person might
be lucky and not come to harm, despite being put at risk. For instance, a young child crossing a busy road may not come to harm,
but we consider preventing this situation part of keeping them safe.}.

\section{Process}\label{sec:process}

The basic idea is that we start with a negated tenet. For example, if we want to ``\node{not harm a person}'', then we begin
with ``\node{harming a person}'' and then consider how this could occur in the current context (i.e.~with respect to the domain
knowledge and system at hand), this is done by asking questions such as: 
\begin{itemize}
\item ``What goals contribute to/against this tenet?''
\item ``How (in this context) could this tenet be violated?'' 
\item ``What does this mean (in this context)?''
\end{itemize}
This process is iterated until all leaf nodes are formalised. The results (the leaf nodes) are negated, yielding a collection of
properties to be checked. It is also possible to use obstacle derivation \textit{a la} vLL to explore the assumptions underlying
the achievement of these properties.

In essence, the process takes a negated tenet, and derives a collection of properties, such that each of the properties, if it
holds, implies the negation of the tenet. Therefore, in order for the desired tenet to hold, the \emph{negation} of each of the
properties must hold. For example, given the tenet that we want to avoid harm to the human, we might derive a collection of
(formally specified) properties that include that we remind the elderly person to eat their lunch, and that we ensure that the
medication they take match what is prescribed.

At a high level, the detailed process is as follows (where the numbers on the left mark different cases, discussed below):
\begin{tabbing}
asd \= asd \= asd \= asd \= asd \= asd \= \kill
\> Initialise root node $N$ with the negation of a tenet \\
\> \textbf{Repeat until} all leaves of refinement tree are formalised \\
\> \> 	Select a leaf node $N$ \\
\> 0. \> 	\textbf{If} $N$ can be formalised \textbf{then} formalise it \\
\> 1. \> \textbf{Elseif} can refine $N$ using domain knowledge  \textbf{then} do so (see Section~\ref{sec:case1})\\
\> 2. \> \textbf{Elseif} can refine $N$ using the goal tree \textbf{then} do so (see Section~\ref{sec:case2})\\
\> 3. \>	\textbf{Else} \\
\> 3a. \> \> \textbf{If} can refine the goal tree by adding relevant goals \\
\> \> \> \> \> \textbf{then} do so (see Section~\ref{sec:case3a})\\
\> 3b. \> \> \textbf{Else} elicit additional relevant domain knowledge (see Section~\ref{sec:case3b})\\
\> \textbf{EndRepeat} \\
\> \textbf{Return} the negations of leaf nodes 
\end{tabbing}

We now consider the different cases.  The ``base case'' is when the node is specific enough that it can be directly formalised
(case 0), in which case we simply formalise the node.  The other cases are: refining using domain knowledge (1), using the goal
tree (2), and expanding either the goal tree (3a) or domain knowledge (3b).  Note that in a situation where a node $N$ can be
refined by applying either domain knowledge or the knowledge implied by the goal-tree, then the process prioritises the domain
knowledge, since it is more general (associated with the problem and the domain), whereas the goal-tree is more specific
(associated with a particular solution).

\subsection{Formalisation (case 0)}\label{sec:case0}

If the node's (informal) description is sufficiently specific that it can be directly formalised, then we refine the node,
replacing it with a node containing an LTL formula. 

Writing a formula may require some attention to the implementation, what information can be monitored, and in what (logical)
form it is provided. For example, in issuing medicine, the implementation (i.e.~the agent's beliefs) may provide predicates such
as $\mathit{issued}(M)$ indicating that medication $M$ was issued, and this predicate would be used in formalising a node
specifying that correct medication was issued. On the other hand, if the belief used was, say,
$\mathit{medicationIssued}(M,P,D,T)$, indicating that medication $M$ was issued to patient $P$ on day $D$ and time $T$, then
this predicate would need to be used instead. 

We assume that formalisation is possible in the following cases in our running example.  Some of the formalisations below could
be improved to be made more precise. For example, instead of $\mathit{emergency} \Rightarrow \mathit{alerted}$ we might require
that should an emergency occur, an alert is sent within a certain time window.

We also assume the obvious compositional property, that a compound formula (e.g.~a conjunction, or negation) can be formalised
exactly when its sub-formulae can be formalised.

\begin{center}
\begin{tabular}{||c|c||}
\hline \hline 
\textbf{Informal Text} & \textbf{Corresponding Formalisation} \\ \hline \hline
remind(X) 
& $\Phi(X)$ 
\\ 
(where $X \in \{ \mbox{breakfast}, \mbox{lunch}, \mbox{dinner}\}$)  &
\\ \hline 
remind(drinkregularly) & $\Psi$ \\ \hline
issued $\neq$ prescribed & $\Diamond (issued(M) \land prescribed(M') \land M \neq M')$ \\ \hline
follow $\lor$ delegate-by-informing & $leave \Rightarrow (\mathit{follow} \lor  \mathit{inform}) $ \\ \hline
{monitor critical incident} & $emergency \Rightarrow alerted$ \\ \hline
{monitor behaviour} & $deteriorated \Rightarrow alerted$ \\ \hline 
out of charge & $\Diamond charge=0$ \\ \hline
obey & $\mathit{request(X)} \Rightarrow \Diamond \mathit{done}(X)$ \\ \hline
\hline
\end{tabular}
\end{center}

Where we define $\Phi(X)$ (when $X$ is a meal), as being true if, at the time of that meal, either the person is eating, or the
system is about to issue a reminder:
$$\Phi(X) \equiv \Box (\mathit{time}(X) \rightarrow (\mathit{eating}(X) \lor \Circle \mathit{remind}(X)))$$

We also define $\Psi$ as being true if the system regularly reminds the person to drink. Specifically, $\Psi$ holds if, whenever
it has been more than 2 hours (``$2h$'') since the last drink, either the system has issued a reminder within the last 15
minutes (``$\bLozenge_{<15m}$''), or the system is about to issue a reminder. In other words, if the person has not had a drink
within the past two hours, then reminders are issued every 15 minutes (until they drink). 
$$\Psi \equiv \Box (\mathit{lastDrink}(T) \land \mathit{now}(T') \land T' > T+ 2h \rightarrow (\bLozenge_{<15m} 
\mathit{remind}(\mathit{drink}) \lor \Circle \mathit{remind}(\mathit{drink})))$$

\subsection{Refining with respect to Domain Knowledge (case 1)}\label{sec:case1}

When refining with respect to domain knowledge, we follow vLL's process: given a domain knowledge rule of the form $\Box (A
\rightarrow P')$, and a node $N$ containing $P$ (where $P$ is in a positive context, and $\exists \theta . P \theta = P'
\theta$), then we refine $N$ with\footnote{The notation $N[A/B]$ denotes the result obtained by taking $N$, and replacing the
sub-expression $A$ with $B$.} $N[P/A\theta]$ (i.e.~replace $P$ with $A \theta$). This gives us the desired relationship between
a node and its child: if $A \rightarrow P$ then $N[P/A] \rightarrow N$.  

Similarly, if $P$ occurs in a negative context (i.e.~within the scope of an odd number of negations), then the rule $\Box (P
\rightarrow A)$ can be applied to replace $P$ with $A$, which gives the desired implication relationship: since $P \rightarrow
A$ we have $(\lnot A) \rightarrow (\lnot P)$ and hence $(\lnot N[P/A]) \rightarrow (\lnot N)$ as desired.  

Note that in either case, if the new node is of the form $\lnot (N_1 \land \ldots \land N_i)$ then we break it into multiple
nodes $\lnot N_i$.  For example, we can use the domain knowledge that keeping someone healthy means (Equation~\ref{eqn6}) that
they have enough food, enough drink, and correct medication. When we refine the node \node{$\lnot$keep healthy} then we replace
it with the node \node{$\lnot (\mbox{enough food} \land \mbox{enough drink} \land \mbox{correct medication})$}, which is then
broken into three nodes.

Note that if the domain knowledge is in the form of a definition ($P' \equiv A$), then, given $P \theta = P' \theta$, we can
replace $P$ with $A\theta$ in any context.  Note that this rule could be applied repeatedly giving a loop, by replacing $A$ with
$P$. We avoid this by assuming that definitions are directional: defining something in terms of something else more specific,
and that the human following the process only goes in the direction of increasing specificity. For example, it would make sense
to refine ``\node{not enough food}'' to ``\node{$<$3 meals a day}'', but not the reverse.

However, what we want to identify as a result of this whole process is not just some ways in which the underlying tenet
(e.g.~``\node{no harm}'') can be violated, but \emph{all} ways in which it can be violated. Therefore, when refining, we
consider not just one domain rule of the form $\Box (X \rightarrow Y)$, but all such rules. 

Specifically, after refining we ask the question: ``is this complete?''. Specifically, given a node $N$, and its refinements
$N_1 \ldots N_i$, the refinements are complete if (following vLL) $(\lnot N_1 \land \ldots \land \lnot N_i) \rightarrow \lnot
N$.  Of course, since nodes being refined have not yet been formalised, this cannot be checked formally, but informally, using
the question\footnote{Note that if the domain knowledge rule used is of the form $A \equiv P'$ then by definition it is
complete.}: ``\textit{if all of these refinements fail to hold, does the original node also necessarily fail to hold?}''. 

For example, when refining ``\node{not kept healthy}'', we refine it into the three sub-nodes: ``\node{not enough food}'',
``\node{not enough drink}'', and ``\node{no/wrong medication}''. We then consider the question: ``\textit{is enough food, enough
drink, and correct medication sufficient to guarantee good health?}''. In this case we might consider that the answer is ``no'',
because health also requires exercise, and attention to psychological well-being (e.g.~companionship, social activities, and a
sense of meaning).  These additional nodes could be added (not shown in Figure~\ref{fig:refinetree}) and further elaborated.

\subsection{Refining with respect to Goals (case 2)}\label{sec:case2}

When refining with respect to goals we derive the process by considering the goal-tree as specifying ``implied'' domain
knowledge. In essence, a goal-subgoal relationship of the form $G_0 \longrightarrow G_1$ (i.e.~$G_1$ is a sub-goal of $G_0$), is
read as implying domain knowledge that bringing about $G_1$ implies $G_0$. In the case where $G_0$ has multiple children, then
when the decomposition is OR each child implies the parent, and when the decomposition is AND then the conjunction of all the
children implies the parent.  

If the (parent) goal $G$ appears in the node $N$ in a positive context (i.e.~it is not in the scope of a negation\footnote{Or,
more precisely, it is in the scope of an \emph{even} number of negations.}), then we
simply use this domain knowledge.  Specifically, we use the implied knowledge of the form $X \rightarrow G$ to create a refined
node $N[G/X]$. There are two sub-cases: if $G$ is OR-refined, then the implied domain knowledge consists of rules $G_i
\rightarrow G$ for each child of $G$ in the goal tree. In this case, we collect the rule applications, giving us multiple
children of $N$, each of the form $N[G/G_i]$.  In the second case, where $G$ is AND-refined in the goal-tree, we create a single
child of $N$: $N[G/\bigwedge_i G_i]$. Note that if $G$ only has a single child, then both sub-cases are equivalent: we add a
refined node $N[G/G_1]$.

For example, given a node \node{keep safe} we can refine it using the goal-tree. Recall (Figure~\ref{fig:goaltree}) that the
goal ``keep safe'' has two (AND-refined) sub-goals (``monitor'' and ``accompany''). Therefore, we have implied domain knowledge
that $(\mbox{monitor} \land \mbox{accompany}) \rightarrow \mbox{keep safe}$. This can be applied, in a similar way to other
domain knowledge, to refine \node{keep safe} to \node{monitor $\land$ accompany}. 

However, what can we do if the node is actually negated, e.g.~\node{$\lnot$keep safe}?  In this case we need to consider
\emph{strengthening} the goal-tree. 

The domain knowledge that we need to refine the node $N$ into more specific sub-nodes is now of the form $\lnot \mathcal{C}
\rightarrow \lnot G$ (where ``$\mathcal{C}$'' denotes an appropriate logical formula combining $G$'s children).  However, this
is not what the goal-tree implies, it implies $\mathcal{C} \rightarrow G$  and hence $\lnot G \rightarrow \lnot \mathcal{C}$. In
order to be able to use the goal-tree in this situation, we need to consider cases where the goal decomposition in the goal tree
is \emph{essential}, that is, where the decomposition of $G$ into $\mathcal{C}$ is such that not only does $\mathcal{C}$ imply
$G$, but it is actually \emph{necessary} not just sufficient, so that we have $G \leftrightarrow \mathcal{C}$.  Then, if $G$
appears in a negative context in the node $N$ then we use the implied knowledge of the form $G \rightarrow \mathcal{C}$. 

There are two cases.  If $G$ is OR-refined, then we have $G \rightarrow (G_1 \lor \ldots \lor G_i)$ and hence
$\lnot (G_1 \lor \ldots \lor G_i) \rightarrow (\lnot G)$ and we replace $G$ (in a negative context in $N$) with $G_1 \lor \ldots
\lor G_i$. The second case is when $G$ is AND-refined, in which case we have $G \rightarrow (G_1 \land \ldots \land G_i)$, hence
$\lnot (G_1 \land \ldots \land G_i) \rightarrow (\lnot G)$ and we replace $G$ (in a negative context in $N$) with $G_1 \land
\ldots \land G_i$, which can be split into multiple sub-nodes $N_i = N[G/G_i]$.

For the example goal-plan tree we assume that this \textbf{strengthened relationship} holds for each of the following nodes and
their children: monitor, keep safe, and $\lnot$harm (also shown as \textbf{bold} labels in shaded nodes in
Figure~\ref{fig:newgoaltree}).  So, for the example we would be able to refine \node{$\lnot$keep safe} into two nodes:
\node{$\lnot$accompany}, and \node{$\lnot$monitor}.

Finally, for the ``$G$ appears in a negative context in $N$'' case, there is one special case to consider, where instead of $N =
\lnot G$, we have $\lnot N = G$, i.e.~instead of $G$ appearing in $N$, actually $N$ appears in $G$. In this case we apply the
same logic as the other negative cases, but, because $G$ doesn't actually appear in $N$ (and $N$ isn't necessarily a negation),
we cannot just replace $N$ with $N[G/\ldots]$. Instead, we replace $N$ with $\lnot \mathcal{G}$ where $\mathcal{G}$ is the
appropriate logical combination of the children of $G$. For example, if $G$ has a single child, $G_1$, then the implied domain
knowledge (assuming a strengthened relationship) is $G \rightarrow G_1$ and hence $(\lnot G_1) \rightarrow (\lnot G)$. We then
have that $N$ is implied by (and hence can be replaced by) $\lnot G_1$, because $(\lnot G_1) \rightarrow (\lnot G)$ and $\lnot G
= \lnot \lnot N \equiv N$.

These cases are summarised in the following table. Note that when refining, we want to make things more specific, so we always
refine a goal in terms of its children, rather than its parent.

\begin{center}
\begin{tabular}{|llll|}
\hline \hline
\textbf{Goal Tree} & \textbf{Implied Domain Knowledge} & $\mathbf{N}$ & $\mathbf{N'}$ \\ \hline \hline
$G \longrightarrow \mathit{OR}(G_1, G_2) $ & $G_i \rightarrow G$ & 
	$N[G]$ & $\{N[G_1], N[G_2]\}$ \\ \hline
$G \longrightarrow \mathit{AND}(G_1, G_2) $ & $(G_1 \land G_2) \rightarrow G$ & 
	$N[G]$ & $N[G_1 \land G_2]$ \\ \hline
$G \longleftrightarrow \mathit{OR}(G_1, G_2) $ & $G \rightarrow (G_1 \lor G_2)$ & & \\
	& $ \; \therefore \; \lnot (G_1 \lor G_2) \rightarrow \lnot G$ & 
	$\lnot N[G]$ & $\lnot N[G_1 \lor G_2]$ \\ \hline
$G \longleftrightarrow \mathit{AND}(G_1, G_2) $ & $G \rightarrow (G_1 \land G_2)$ & & \\
& $  \; \therefore \; \lnot (G_1 \land G_2) \rightarrow \lnot G$ & &  \\
	& $  \; \therefore \; ((\lnot G_1) \lor (\lnot G_2)) \rightarrow \lnot G$ & 
	$\lnot N[G]$ & $\{ \lnot N[G_1], \lnot N[G_2]\}$ \\ \hline
$G \longleftrightarrow \mathit{OR}(G_1, G_2) $ & $G \rightarrow (G_1 \lor G_2)$ & \multicolumn{2}{c|}{(where $G = \lnot N$)} \\
	& $ \; \therefore \; \lnot (G_1 \lor G_2) \rightarrow \lnot G$ & 
	$N$ & $\lnot (G_1 \lor G_2)$ \\ \hline
$G \longleftrightarrow \mathit{AND}(G_1, G_2) $ & $G \rightarrow (G_1 \land G_2)$ & \multicolumn{2}{c|}{(where $G = \lnot N$)} \\
& $  \; \therefore \; \lnot (G_1 \land G_2) \rightarrow \lnot G$ & &  \\
	& $  \; \therefore \; ((\lnot G_1) \lor (\lnot G_2)) \rightarrow \lnot G$ & 
	$N$ & $\{ \lnot G_1, \lnot G_2\}$ \\ 	
\hline \hline
\end{tabular}
\end{center}

\subsection{Expanding the Goal Tree (case 3a)}\label{sec:case3a}

There is one slight complication when refining with respect to goals. This is a situation where the goal-tree is ``missing'' a
goal. That is, where instead of having $G \longrightarrow G_1$ ($G_1$ has $G$ as a parent), $G_1$ has a different, more
general, parent goal. For instance, in the goal-tree of Figure~\ref{fig:goaltree} the two goals ``\node{Keep Healthy}'' and
``\node{Keep Safe}'' have as their parent the (more general) goal ``\node{Support}'', rather than a more specific goal
``\node{Keep from harm}''. This means that if the refinement tree has a node relating to harming someone, then we cannot use the
goal tree to refine the node. 

In this situation, where we might want to use the child goals to refine nodes, we may need to conceptually consider additional
``phantom'' nodes in the goal-tree. This is done by adding intermediate nodes to the tree, as illustrated below.

\begin{center}
\includegraphics[width=\textwidth]{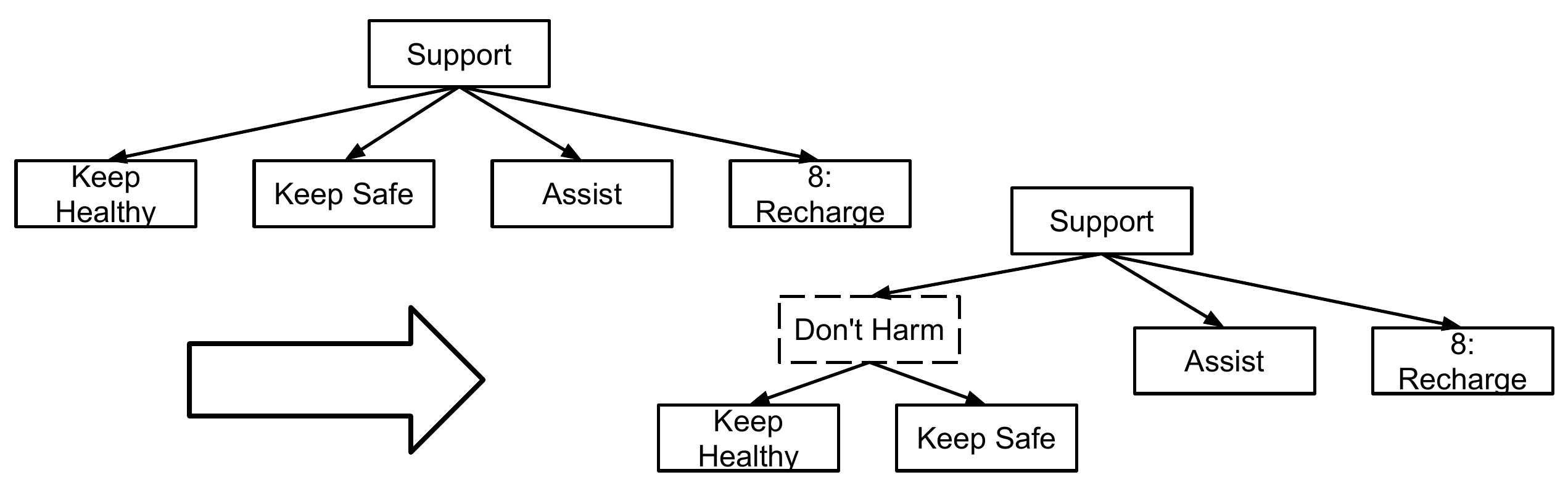}
\end{center}

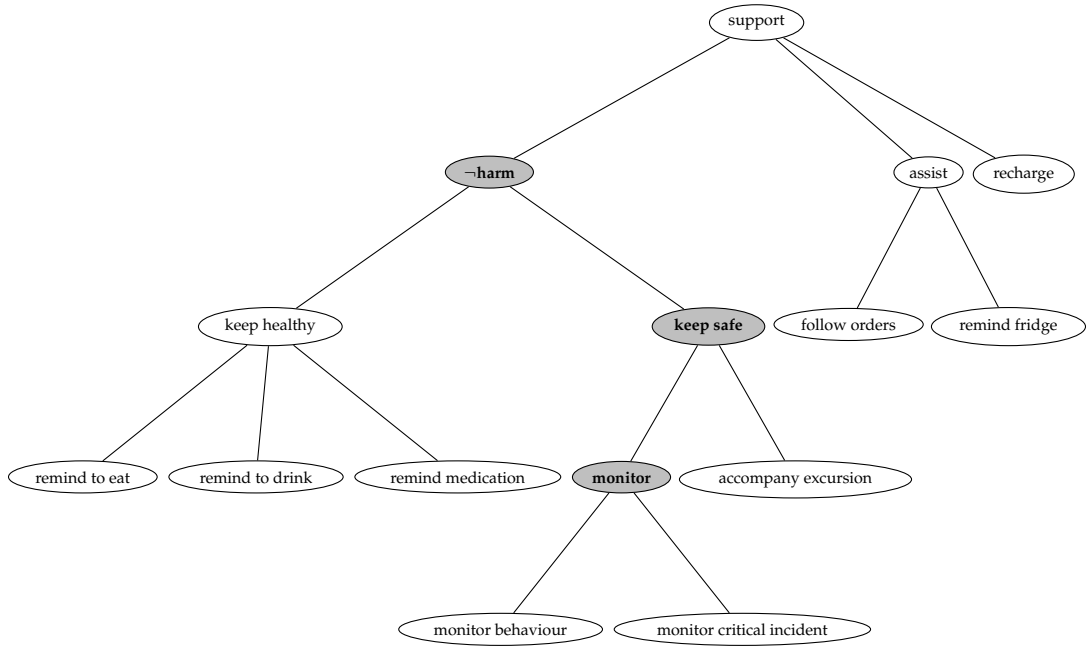
\begin{figure}[p]
\centering
\tiny 
\begin{forest} for tree={ellipse,draw,l=2cm},
[{support} [{\textbf{$\lnot$harm}},fill=lightgray [{keep healthy} [{remind to eat}  ][{remind to drink}  ][{remind medication}  ] ][{\textbf{keep safe}},fill=lightgray [{\textbf{monitor}},fill=lightgray [{monitor behaviour}  ][{monitor critical incident}  ] ][{accompany excursion}  ] ] ][{assist} [{follow orders}  ][{remind fridge}  ] ][{recharge}  ] ]
\end{forest}
\caption{Revised Example Goal Tree (bold shaded nodes indicate ones with strengthened relationship to their children)}\label{fig:newgoaltree}
\end{figure}

The resulting goal-tree is shown in Figure~\ref{fig:newgoaltree} (on page \pageref{fig:newgoaltree}). It adds an intermediate
node  ``$\lnot$harm''.

\subsection{Expanding the Domain Knowledge (case 3b)}\label{sec:case3b}

Finally, if neither case applies, then we can proceed by expanding domain knowledge. This is done by considering the node and
asking ``how can this occur?'', or ``what does this mean?'' (specifically: ``what counts as this node?''). The new domain
knowledge can then be used to continue the process.

\section{Results}

\begin{figure}[htp]
\tiny
\begin{forest} for tree={draw,grow'=0,l=2cm,anchor=west,child anchor=west},
[{harm:g} 
[{$\lnot$keep healthy:d6} 
[{$\lnot$enough food:d3} 
[{$<$3 meals a day:d9} 
[$\lnot$do(breakfast)\\$\land$do(lunch)\\$\land$do(dinner):d2,align=center,base=bottom 
[{$\lnot$do(breakfast):d2} 
[{$\lnot$remind(breakfast):f} 
[{$\lnot$$\Phi_{\mathit{breakfast}}$} ] ] ] 
[{$\lnot$do(lunch):d2} 
[{$\lnot$remind(lunch):f} 
[{$\lnot$$\Phi_{\mathit{lunch}}$}  ] ] ] 
[{$\lnot$do(dinner):d2} 
[{$\lnot$remind(dinner):f} 
[{$\lnot$$\Phi_{\mathit{dinner}}$}  ] ] ] ] ] ] 
[{$\lnot$enough drink:d4} 
[{$<$1.2L/day:d7} 
[{$\lnot$do(drinkregularly):d2} 
[{$\lnot$remind(drinkregularly):f} 
[{$\lnot$$\Psi$}  ] ] ] ] ] 
[{$\lnot$correct medication:d8} 
[{issued $\neq$ prescribed:f} 
[{$\Diamond$(issued(m)$\land$prescribed(mp)$\land$neq(m,mp))}  ] ] ] ] 
[{$\lnot$keep safe:g} 
[{$\lnot$monitor:g} 
[{$\lnot$monitor behaviour:f} 
[{$\lnot$ (deteriorated$\Rightarrow$alerted)}  ] ] 
[{$\lnot$monitor critical incident:f} 
[{$\lnot$ (emergency$\Rightarrow$alerted)}  ] ] ]
[{$\lnot$accompany excursion:d5} 
[{$\lnot$follow or delegate-by-informing:f} 
[{$\lnot$ (leave$\Rightarrow$follow$\lor$inform)}  ] ] ] ] ]
\end{forest}
\caption{Refinement Tree for Tenet ``do not harm''}\label{fig:refinetree}
\end{figure}
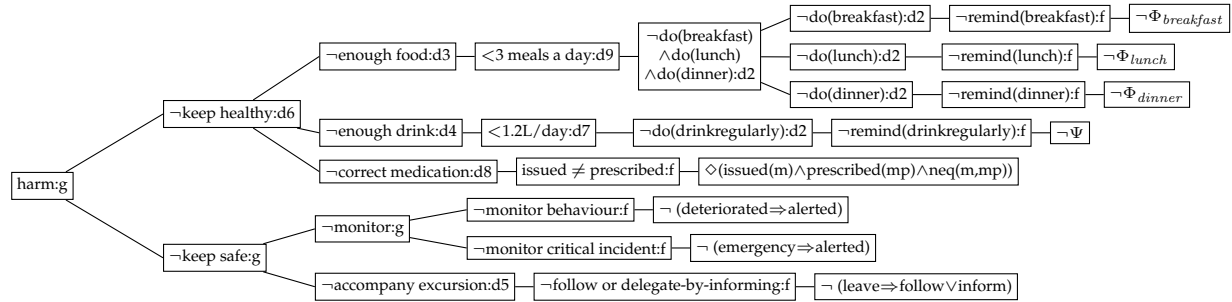

For the running example, starting with the high-level tenet ``do not harm'', the given goal tree, and the domain  knowledge,
following the process might\footnote{Since the process is a design process, followed by humans, it is not deterministic.} yield
the refinement tree shown in Figure~\ref{fig:refinetree}.  Annotations in the tree of the form ``d$n$'' indicate that the node
was refined using domain knowledge rule $n$. An annotation of ``g'' indicates that the goal tree was used, and ``f'' indicates
that the node was able to be formalised.

This figure shows a process that began with the root node ``harm'' (the negated tenet), and was then refined by considering the
goal-tree, where the goals ``keep healthy'' and ``keep safe'' both interfere with the negated tenet (and where the goal tree had
been refined to create a ``do not harm'' parent to these goals).

Each of these goals is then refined by eliciting additional domain knowledge and applying that. Specifically, that being healthy
involves having enough to eat, enough to drink, and taking medication; and that being safe involves ensuring adequate monitoring
when leaving the house, and adequate monitoring to detect any health emergencies or accidents. 

We then further refine these nodes with (existing) domain knowledge. For instance, that ``not enough food'' means fewer than
three meals a day. We also use domain knowledge that ensuring (e.g.) three meals a day can be achieved by issuing appropriate
reminders. Then, since this is sufficiently specific, we formalise this.  Similarly, we can refine not having enough drink to
drinking regularly, and formalise it in terms of reminders.

Then we can take the negated leaf nodes of the refinement tree, which gives us the following formulae for verification:
$$\Phi({\mathit{breakfast}}), \;  \Phi({\mathit{lunch}}),  \; \Phi({\mathit{dinner}}) $$
$$ \Box (\mathit{lastDrink}(T) \land \mathit{now}(T') \land T' > T+ 2h \rightarrow (\bLozenge_{<15m} 
\mathit{remind}(\mathit{drink}) \lor \Circle \mathit{remind}(\mathit{drink})))$$
$$\Box (\mathit{issued}(M) \land \mathit{prescribed}(M')) \rightarrow M = M'$$
$$\Box (\mathit{leave} \rightarrow (\mathit{follow} \lor \mathit{inform})) $$
$$\Box \mathit{emergency} \rightarrow \mathit{alerted}$$
$$\Box \mathit{deteriorated} \rightarrow \mathit{alerted}$$
where
$\Phi(X) \equiv \Box (\mathit{time}(X) \rightarrow (\mathit{eating}(X) \lor \Circle \mathit{remind}(X)))$

\section{Next Steps}

This working note presented a high-level process that can be used by a human designer to systematically identify a set of
verification properties. 

However, there is still much work to be done. In particular, we need to find answers to the following questions:
\begin{itemize}

\item Can this process be used in practice by designers (other than the author of this paper)? 

\item Can this process be used for a wider range of examples, including larger and more complicated ones?

\item How effective is this process at identifying important properties to be verified?

\item How can this process be supported by tools?

\end{itemize}

Other areas for future work include: richer representation (e.g.~logic with probabilities, other design models, graphical model
for domain knowledge); priorities between tenets, perhaps linking to ideas on human values (e.g.~\cite{Schwartz2012}); and
dealing with conflict between goals, for instance in the running example we want the system to obey the user, but what if the
user asks the system to stop issuing reminder to drink?


\end{document}